# Redesigning Photonic Interconnects with Silicon-on-Sapphire Device Platform for Ultra-Low-Energy On-Chip Communication

Venkata Sai Praneeth Karempudi, Sairam Sri Vatsavai, Ishan Thakkar
Department of Electrical and Computer Engineering, University of Kentucky, Lexington, Kentucky, USA.
{kvspraneeth, Sairam_Srivatsavai, igthakkar}@uky.edu

## ABSTRACT

Traditional silicon-on-insulator (SOI) platform based on-chip photonic interconnects have limited energy-bandwidth scalability due to the optical non-linearity induced power constraints of the constituent photonic devices. In this paper, we propose to break this scalability barrier using a new silicon-on-sapphire (SOS) based photonic device platform. Our physical-layer characterization results show that SOS-based photonic devices have negligible optical non-linearity effects in the mid-infrared region near 4µm, which drastically alleviates their power constraints. Our link-level analysis shows that SOS-based photonic devices can be used to realize photonic links with aggregated data rate of more than 1 Tb/s, which recently has been deemed unattainable for the SOI-based photonic on-chip links. We also show that such high-throughput SOS-based photonic links can significantly improve the energy-efficiency of on-chip photonic communication architectures. Our system-level analysis results position SOS-based photonic interconnects to pave the way for realizing ultra-low-energy (< 1 pJ/bit) on-chip data transfers.

## 1. INTRODUCTION

With rapidly increasing demand for data-centric high-performance computing, future manycore processors will require exceedingly high communication bandwidth from the on-chip interconnection networks. However, traditional electrical networks-on-chip (ENoCs) already consume extravagantly large amount of chip area and total system power, which makes the energy-efficient scaling of their bandwidth improbable. This motivates the need for a new interconnect technology that can be leveraged to realize extremely high-throughput (>1 terabits/s) and energy-efficient (<1 pJ/bit) interconnects for future manycore computing architectures.

Recent advancements in CMOS-photonics integration [1] have enabled an exciting solution in the form of photonic NoCs (PNoCs). Several PNoC architectures have been proposed thus far (e.g., [2]-[4]). PNoC architectures typically employ on-chip photonic links, each of which connects two or more clusters of processing cores. Each photonic link comprises of one or more photonic silicon-on-insulator (SOI) waveguides with dense wavelength division multiplexing (DWDM) of multiple wavelengths into each waveguide. In a DWDM-enabled SOI waveguide, SOI microring resonator (MR) modulators, which are arrayed along the waveguide at the source end, modulate input electric signals onto parallel photonic channels. The photonic channels travel through the waveguide and reach the destination end, where an array of SOI MRs drop the parallel photonic signals onto the adjacent photodetectors to recover the electric data signals. Thus, DWDM that utilizes SOI photonic devices enables high-bandwidth parallel data transfer in PNoCs.

A critical parameter for designing a high-throughput SOI-based photonic (*SOIPh*) link is its optical power budget (OPB), which determines the upper limit of the allowable signal losses and power penalties in the link for the given aggregated data rate (#DWDM channels ($N_\lambda$) × channel bitrate) of the link. The OPB of a *SOIPh* link is the difference between the photodetector noise floor (i.e., detector sensitivity which has a dependency on bit-rate [16]) and the maximum allowable optical power (MAOP) in the link. The MAOP in a *SOIPh* link is determined by the optical non-linear effects of silicon in the constituent *SOIPh* waveguides and MR modulators. The primary non-linear effect in silicon at the operating wavelengths of the *SOIPh* platform (i.e., 1.3µm-1.6µm) is two-photon absorption (TPA) [17], which has been shown to induce strong free carrier absorption (FCA) and free-carrier dispersion (FCD) effects in silicon [29], significantly increasing the absorption losses in *SOIPh* waveguides [23] and causing self-heating and irreparable resonance shifts in *SOIPh* MR modulators [17]. As recently demonstrated in [16], these TPA induced effects limit the MAOP in *SOIPh* links below 20dBm, which in turn restricts the achievable link data rate below 900 Gb/s and energy-efficiency above ~2 pJ/bit, even for the most optimistic *SOIPh* device parameters from [8]. Therefore, to achieve >1 terabits/s aggregated data rate and sub-pJ/bit energy-efficiency for *SOIPh* links, which is a very important step towards realizing the exascale computing systems of the future [37], *the TPA effect in silicon must be alleviated to increase the MAOP in SOIPh links*.

In this paper, we present silicon-on-sapphire (SOS) based photonic platform as a potential solution that can mitigate the TPA related shortcomings of the *SOIPh* platform. The fact that underpins our *rationale* is that SOS-based photonic (*SOSPh*) waveguides and MRs have been shown to exhibit low absorption losses and no TPA for the operating wavelengths in the mid-infrared region near 4µm [19][30]. The *SOSPh* platform has these advantages near 4µm wavelength region, compared to the *SOIPh* platform, because near 4µm wavelength sapphire has lower material losses than $SiO_2$ [31] and silicon bandgap is smaller than the total energy of two absorbed photons [32]. Although several prior works have demonstrated the usefulness of *SOSPh* devices for optical signal processing (e.g., [25]-[29]), *no prior work has yet explored SOSPh devices for realizing on-chip interconnects*. Therefore, in this paper, using our detailed modeling at the device- and link-level as well as extensive system-level analysis, *we show for the first time* that *SOSPh* interconnects can pave the way for realizing extreme-throughput (>1 terabits/s) and ultra-low-energy (<1 pJ/bit) on-chip data communication. The key contributions of this paper are summarized below:

- We characterize different types of losses and optical properties of *SOSPh* waveguides and MRs to derive compact design models;
- We use our developed compact models to derive a new set of guidelines for designing *SOSPh* links and PNoC architectures;
- We utilize our developed guidelines to optimize the designs of *SOSPh* links, and then compare their aggregated data rate and energy-per-bit values to the optimized designs of *SOIPh* links;
- We evaluate the impact of optimized designs of *SOSPh* and *SOIPh* links on the performance and energy-efficiency of a well-known Clos [33] PNoC architecture.

## 2. MOTIVATION

To demonstrate the limitations of the *SOIPh* device platforms in general, we used different *SOIPh* platforms from [6], [34]-[36] to perform a design analysis for on-chip links following the more realistic design guidelines given in [16]. Results of our analysis are given in Fig. 1. Fig. 1(a) depicts how the OPB in various *SOIPh* links (corresponding to the *SOIPh* platforms from [6], [34]-[36]) is utilized depending on the losses present in the links, whereas Fig. 1(b) shows the best achievable aggregate data rate (i.e., #DWDM channels ($N_\lambda$) × channel bitrate) and energy-per-bit (EPB) values for the links. We also show our projected results for our target (preferred) photonic platform. In Fig. 1(a), the MAOP for the OPB values of all *SOIPh* links is considered to be 20dBm. Moreover, the EPB values in Fig. 1(b) present total EPB values that include contributions from the link laser power, thermal tuning power, modulator driver power, and receiver power, as outlined in the guidelines from [20]. From Fig. 1(a), different *SOIPh* links experience different amounts of total optical power loss (including crosstalk and signal degradation related power penalties [18]). This total power loss whittles down the OPB of all *SOIPh* links, leaving only a smaller portion of the OPB available to support aggregated data rate. For example, among all considered *SOIPh* platforms, the *SOIPh* platform named "zero-change" from [6] has the largest OPB of 51.5dB, which corresponds to -31.5dBm detector sensitivity and the TPA-limited MAOP of 20dBm [23]. From this 51.5dB OPB, 21.15dB portion is whittled down due to optical losses, which leaves 30.35dB of the OPB available for supporting the highest data rate in Fig. 1(b) of 636 Gb/s. This larger value of aggregate data rate better prorates the EPB contributions from laser power, thermal tuning, modulator power, and receiver power to yield the lowest total EPB value in Fig. 1(b) of 2.1pJ//bit for the *SOIPh* platform "zero-change".

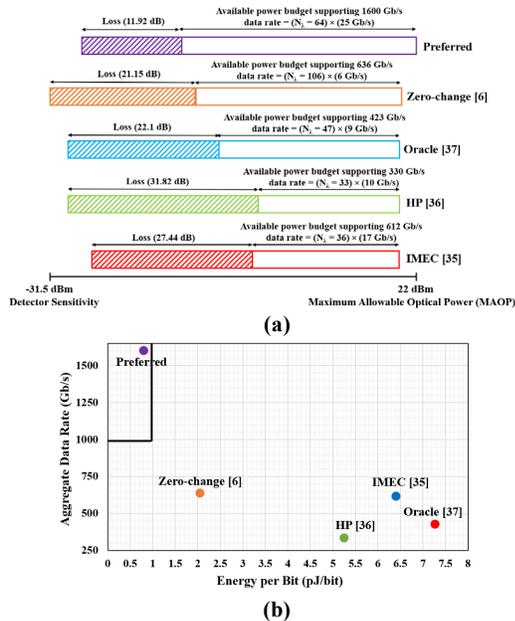

Fig. 1: (a) Distribution of optical power budget (OPB), and (b) Best achievable aggregate data rate (#DWDM channels ($N_\lambda$) × channel bitrate) vs energy-per-bit (EPB), for our analyzed photonic links based on the *SOIPh* platforms from prior works and our target (preferred) photonic platform.

Clearly, higher aggregate data rate and lower EPB can be achieved for the "zero-change" platform, if the MAOP for it can be increased from 20dBm and/or total optical loss can be decreased, so that a larger portion of its OPB can be rendered available to support larger aggregate date rate (i.e., larger $N_\lambda$ and/or higher channel bitrate). Therefore, we envision a target platform (Fig. 1) that can increase the MOAP to 22dBm and reduce the total optical loss to 11.9dB, to yield a higher OPB that can support aggregate data rate of up to 1600 Gb/s and EPB of <1pJ/bit. The cross-layer modeling and analysis results presented in the rest of this paper show how our envisioned target platform for the design of extreme-throughput (>1 terabits/s) and ultra-low-energy (<1 pJ/bit) photonic interconnects can be realized using our proposed *SOSPh* devices and links.

## 3. MODELING OF SOS-BASED DEVICES

It is established from prior works (e.g., [16][40]) that the performance and energy-efficiency of photonic interconnects depend on the optical characteristics of the constituent waveguides and MR devices. Crucial optical characteristics for photonic interconnects include optical losses in waveguides and spectral footprints (e.g., Q-factor, free-spectral range (FSR)) of MR devices. In this section, we derive compact models for the optical characteristics of *SOSPh* waveguides and active/passive MR devices, and compare these models with the compact models for *SOIPh* devices from prior work. As *SOSPh* devices have been shown to exhibit low absorption losses and no TPA for wavelengths near 4μm region [19][30], we model the *SOSPh* devices to be operating at wavelengths near 4μm.

### 3.1 Modeling of SOS-Based Passive Devices
*Modeling of SOS-based passive waveguides:*
We use Fourier and finite difference time domain (FDTD) analysis methods using a commercial grade tool from Lumerical [44], to model the dimensions and losses in SOS passive waveguides. From our analysis, the cross-sectional dimensions of an SOS channel waveguide (Fig. 2(a)) that can support the single-mode operation near 4μm wavelength with at least 80% optical confinement were found to be 1200nm×970nm, which are significantly larger than the dimensions (450nm×220nm) of a typical SOI channel waveguide operating near 1.5μm. We evaluate the scattering loss and absorption loss of SOS and SOI channel waveguides using the models and methods from [7] and [19]. Our evaluated loss values are given in Table 1. Both silicon and sapphire exhibit lower material loss near 4μm region [32], which results in lower absorption loss for SOS waveguides. On the other hand, from [38], the scattering loss in a waveguide depends on the core-cladding refractive-index contrast (Δn) and the ratio (σ/λ) of waveguide sidewall roughness (σ) to the operating wavelength (λ). With negligible differences in Δn and σ between SOI and SOS waveguides (Table 1), longer operating wavelengths results in lower scattering losses for SOS waveguides.

Table 1: Various types of losses and optical parameters for *SOSPh* and *SOIPh* devices.

| Type of Loss | SOS | SOI |
|---|---|---|
| Waveguide Scattering Loss (dB/cm) | 0.9374 | 1.4 |
| Waveguide Absorption Loss (dB/cm) | $10^{-8}$ | 0.1 |
| Waveguide sidewall roughness (σ) (nm) [7] | 4 | 6 |
| Core-cladding refractive-index contrast (Δn) | 1.67 | 2.06 |
| MR Bending Loss (dB/rad) | 0.004 | 0.0073 |

*Modeling of SOS-based passive MRs:*
In this subsection, we present our compact models that relate an MR's Q-factor with its radius (R) and coupling gap size (g) (i.e., gap size between the rectilinear waveguide and MR) (Fig. 2(a)). From [9], the Q-factor of an MR depends on the total round-trip loss in the MR's waveguide, which is the sum of scattering loss, absorption loss and bending loss (Table). To derive the bending loss values (Table 1), we used the eigenmode-solver based methods described in [39].

As a first step towards deriving our intended compact models, we analyzed coupling coefficient (κ) of an MR as a function of R and g. As g increases, the power coupled into the MR from the rectilinear waveguide decreases, which in turn decreases κ. For SOS and SOI MRs, κ can be calculated using Eq. (1) [8]:

$$\kappa = \sin\left(2\pi \frac{L}{\lambda_{res}} \frac{n_{\text{eff,even}} - n_{\text{eff,odd}}}{2}\right) \qquad (1)$$

Where L is the MR circumference given as L = 2 × π × (MR radius (R)), $\lambda_{res}$ is the MR's resonance wavelength, $n_{\text{eff,even}}$ is even-mode effective index and $n_{eff,odd}$ is the odd-mode effective index. We used FDTD simulations to extract $n_{eff,even}$ and $n_{eff,odd}$ values.

Fig. 2 gives κ values for *SOSPh* and *SOIPh* MRs as a function of *g* and R. From the figure, for R = 10µm and *g* = 50nm, κ = 0.987 for the *SOSPh* MR, whereas κ = 0.92 for the *SOIPh* MR. Thus, *SOSPh* MRs achieve larger values of κ at lower gap sizes. Also, for = 15µm, as *g* increases from 50nm to 150nm, κ for *SOSPh* MRs decreases from 0.988 to 0.4825, whereas for *SOIPh* MRs κ decreases from 0.92 to 0.39. Thus, for *SOSPh* MRs κ decreases less rapidly with increase in *g* compared to *SOIPh* MRs.

This type of intricate behavior of κ results into an elaborate relation of MR Q-factor with R and *g*. To characterize this relation, we plugged our obtained κ values from Table 1 in Eq. (2) [9]:

$$Q = \frac{\pi\, n_g\, L\sqrt{ra}}{\lambda_{res}\,(1 - ra)} \quad (2)$$

Where $n_g$ is group index of silicon, r is cross coupling coefficient (r = $\sqrt{1 - \kappa^2}$), a is round-trip loss coefficient, with other symbols defined with Eq. (1). To obtain a, total loss for a round trip length of an MR along its circumference L is calculated based on the loss values from Table 1. Our obtained Q-factor values for *SOSPh* and SOIPh MRs are shown in Fig. 3(a) and 3(b).

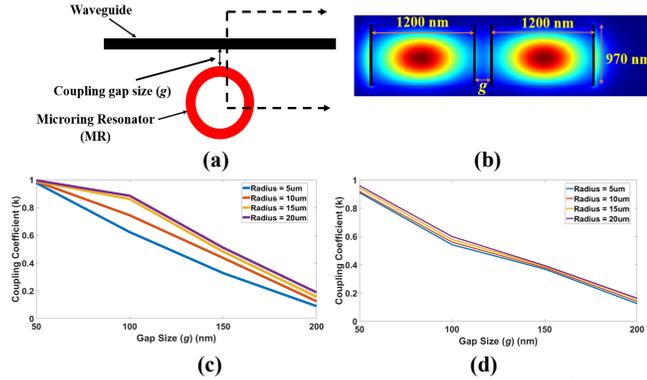

(a) (b) (c) (d)

**Fig. 2:** (a), (b) Depiction of the cross-sectional dimensions of and the coupling gap size (*g*) between a waveguide and an MR; and MR coupling coefficient (κ) as a function of gap size (*g*) and MR radius (R) for (c) *SOSPh* platform and (d) *SOIPh* platform.

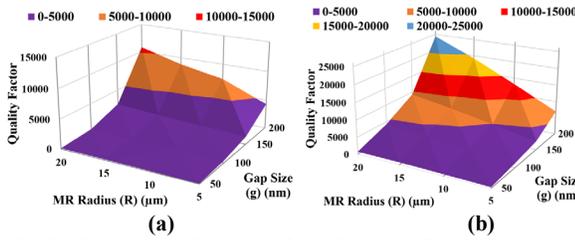

(a) (b)

**Fig. 3:** Quality factor (Q-factor) based on coupling gap size (*g*) and MR radius (R) for (a) SOSPh platform, and (b) SOIPh platform.

From Fig. 3, for given R and *g* values, Q-factor values for SOS MRs are lower compared to SOI MRs. This is because, r is lower for SOS MRs compared to SOI MRs (e.g., for R=10µm and *g*=50nm, r=0.67 for SOS MRs, where it is 0.87 for SOI MRs), which together with longer operating wavelengths for SOS MRs (i.e., λres ≈ 4µm) results in lower Q-factor values for SOS MRs.

### 3.2 Modeling of SOS-Based Active MRs

Active tuning of MRs' resonance wavelengths is required not only for realizing active devices such as modulators and switches [15], but also for counteracting the fabrication process and thermal variations induced unwanted resonant shifts [42]. A common method of achieving active resonance tuning in MRs is to change the free-carrier concentration in MR cores [15], which in turn changes the MR core's (which is made of silicon in both *SOSPh* and *SOIPh* platforms) refractive index (Δn) and absorption loss coefficient (Δα) due to the free-carrier dispersion (FCD) and free-carrier absorption (FCA) effects in silicon [41]. We model the relation of Δn and Δα with the change in free-carrier concentration using the following equations [14]:

*FCD-FCA Equations for SOS (operating wavelength of ~4µm):*
$$\Delta\alpha = (7.45 \times 10^{-22}\Delta N_e^{1.245} + 5.43 \times 10^{-20}\Delta N_h^{1.153}) \quad (3)$$
$$\Delta n = -(7.25 \times 10^{-21}\Delta N_e^{0.991} + 9.99 \times 10^{-18}\Delta N_h^{0.839}) \quad (4)$$

*FCD-FCA Equations for SOI (operating wavelength of 1.55µm):*
$$\Delta\alpha = (3.0 \times 10^{-18}\Delta N_e + 2.0 \times 10^{-18}\Delta N_h) \quad (5)$$
$$\Delta n = -(6.2 \times 10^{-22}\Delta N_e + 6.0 \times 10^{-18}\Delta N_h^{0.8}) \quad (6)$$

Where $\Delta N_e$ is free-electron concentration and $\Delta N_h$ is free-hole concentration. For given $\Delta N_e = 10^{17}$cm$^{-3}$ and $\Delta N_h = 10^{18}$ cm$^{-3}$, Δα and absolute Δn values are higher for SOS MRs (i.e., Δα = 4.21, |Δn| = 13.1×10$^{-3}$ compared to SOI MRs (i.e., Δα = 2.3, |Δn| = 1.56×10$^{-3}$), *which means that active tuning of SOS MRs can be achieved with greater energy-efficiency*. To evaluate the energy-efficiency of active tuning, we model the dynamic energy-per-bit for tuning ($E_{tuning}$) of SOS/SOI MRs with the following equation [15]:

$$E_{tuning} = \frac{V}{4}\,\frac{n_g q\, J}{\lambda_r n_f \Gamma}\,\Delta\lambda_m \quad (7)$$

Where V is the tuning voltage across the MR core required to effect the desired change in free-carrier concentration inside the MR core, $n_g$ is group index of silicon, q is charge of an electron, **J** is the bulk volume of the MR core in which the change in free-carrier concentration occur, $\lambda_{res}$ is MR resonance wavelength, Γ is the mode confinement factor (typically Γ = 0.8), $n_f$ is the ratio of Δn for silicon to the electron-hole pair density that can be evaluated using the formula give in [43] (e.g., $n_f = 2.3\times10^{-20}$cm$^3$ for SOS and $n_f = 2.13\times10^{-21}$cm$^3$ for SOI [43]) and $\Delta\lambda_m$ is the magnitude of wavelength tuning.

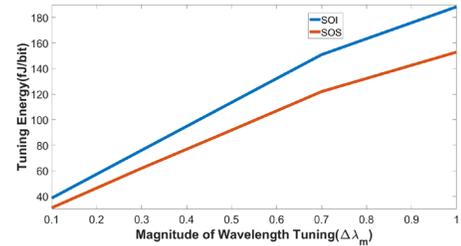

**Fig. 4:** MR tuning energy versus magnitude of wavelength tuning ($\Delta\lambda_m$) for SOSPh and SOIPh MRs.

Fig. 4 shows $E_{tuning}$ as a function of $\Delta\lambda_m$. From the figure, $E_{tuning}$ for SOS MRs is lower than that for SOI MRs for the entire range of $\Delta\lambda_m$, which corroborates our earlier observation that the active tuning of SOS MRs can achieve greater energy-efficiency.

Using the device-level compact models derived in this section, we develop new physical-layer design guidelines for *SOIPh* and *SOSPh* on-chip links, as described in the next section. Using these guidelines, we evaluate the achievable aggregated datarate and energy-per-bit values for *SOIPh* and *SOSPh* on-chip links.

### 4. LINK-LEVEL MODELING AND ANALYSIS

From [16], the achievable aggregated data rate and energy-per-bit (EPB) values for photonic links not only depend on the OPB of the links and optical characteristics of the constituent devices, but also on several physical-layer design parameters such as the number of DWDM wavelengths ($N_\lambda$), free-spectral range (FSR), and OPB. For designing a photonic link, $N_\lambda$ is the most important design parameter and OPB is the most critical design constraint. For a link, to find the best value of $N_\lambda$ that can optimally utilize its OPB, the condition given in Eq. (8) should be satisfied.

$$\text{OPB (dB)} \geq P_{loss}^{dB} + 10\log_{10}(N_\lambda) \quad (8)$$

$$OPB\ (dB) = MAOP - detector\ sensitivity \quad (9)$$

$P_{loss}^{dB}$ in Eq. (8) accounts for total losses in the link including the signal truncation penalty and modulator/detector crosstalk penalty [11]. From [11], the crosstalk and signal truncation penalties depend on MR Q-factor, channel bit-rate, and inter-channel spacing (which relates to FSR and $N_\lambda$ [12]). Moreover, the detector sensitivity in Eq. (9) also depends on channel bit-rate [16]. Therefore, for given values of MR Q-factor, FSR, and MAOP (Eq. (9)), only a unique combination of $N_\lambda$ and bit-rate can optimally utilize the available OPB while satisfying the condition in Eq. (8). This unique optimal combination of $N_\lambda$ and bit-rate determines the best achievable aggregate data rate (i.e., $N_\lambda$ × bit-rate) and energy-per-bit (EPB) for the link [16].

To evaluate the impacts of SOS and SOI devices on the data rate and EPB of links, we use the guidelines given in [16] (as done for our analysis in Section 2) to design *SOIPh* and *SOSPh* on-chip links for four different combinations of MR Q-factor, FSR, and MAOP shown in Table 2. For *SOIPh* links, we choose the TPA-limited MAOP value of 20dBm [23]. In contrast, due to the absence of TPA in *SOSPh* links, it is intuitive to consider a very high value of MAOP. However, we consider a conservative MAOP value of 22dBm for *SOSPh* links. Our rationale for being conservative is that a not-too-high value of MAOP is more likely to require a reasonable amount of per-wavelength optical power. In contrast, a very high value of MAOP (e.g., >25dBm) can require per-wavelength optical power of greater than 5dBm, which might be very difficult to extract from the state-of-the-art comb laser sources [18]. Moreover, in Table 2, we choose the MR Q-factor values in the range from 6000-9000, as it is shown in [11] that this range of Q-factor values can yield minimal values of signal truncation and crosstalk penalties. For these Q-factor values in Table 2, we use the device-level compact models from Section 3 to reckon the corresponding values of MR radius R, which we use in Eq. (10) to reckon the corresponding FSR values.

$$\text{FSR} = \frac{\lambda_{res}^2}{2\pi R n_g} \quad (10)$$

Where $\lambda_{res}$ is MR's resonance wavelength and $n_g$ is MR's group index, which we evaluate using finite difference element method in Lumerical's MODE tool [44].

We use the values from Table 2 to design *SOSPh* and *SOIPh* links for a well-known PNoC architecture: a 256-core 8-ary 3-stage CLOS PNoC [33]. We consider the worst-case link of CLOS PNoC that has the length of 4.5cm for 22nm technology node [33]. Then, for each value combination in Table 2, we sweep the bit-rate in the range from 1Gb/s to 40 Gb/s, and use the exhaustive search based heuristic from [12] to find the optimal $N_\lambda$ for each considered bit-rate value. Then, for each considered bit-rate value, we evaluate aggregate data rate ($N_\lambda$ × bit-rate) and total EPB (laser + thermal tuning + modulator driver + receiver) values using EPB models from [16]. These evaluated data rate and EPB values are plotted in Fig. 5.

**Table 2: Considered Q-factor, FSR, and MAOP values for our analyzed SOSPh and SOIPh links.**

|  | **Considered Q-factor, FSR, and MAOP Values** |
|---|---|
| **SOSPh Links** | Q-factor=6000, FSR=80nm, MAOP=22dBm |
|  | Q-factor=7000, FSR=60nm, MAOP=22dBm |
|  | Q-factor=8000, FSR=48nm, MAOP=22dBm |
|  | Q-factor=9000, FSR=40nm, MAOP=22dBm |
| **SOIPh Links** | Q-factor=6000, FSR=20nm, MAOP=20dBm |
|  | Q-factor=7000, FSR=15nm, MAOP=20dBm |
|  | Q-factor=8000, FSR=13nm, MAOP=20dBm |
|  | Q-factor=9000, FSR=11nm, MAOP=20dBm |

Fig. 5 (a) (Fig. 5(b)) shows the aggregate data rate and EPB values for four different *SOSPh* (*SOIPh*) links that correspond to the four combinations of Q-factor, FSR, and MAOP values from Table 2. From the figures, the peak aggregate data rate values for four *SOSPh* links are 1600 Gb/s, 1350 Gb/s, 1200 Gb/s and 1100 Gb/s, and their corresponding EPB values are 1.15 pJ/bit, 1.14 pJ/bit, 1.13 pJ/bit and 1.12 pJ/bit, respectively. On the other hand, the peak aggregate data rate values for *SOIPh* links are 697 Gb/s, 630 Gb/s, 612 Gb/s and 590 Gb/s, and their corresponding EPB values are 2.09 pJ/bit, 2.22 pJ/bit, 2.23 pJ/bit and 2.28 pJ/bit, respectively. Clearly, *SOSPh* links achieve higher aggregate data rate and lower EPB values compared to *SOIPh* links.

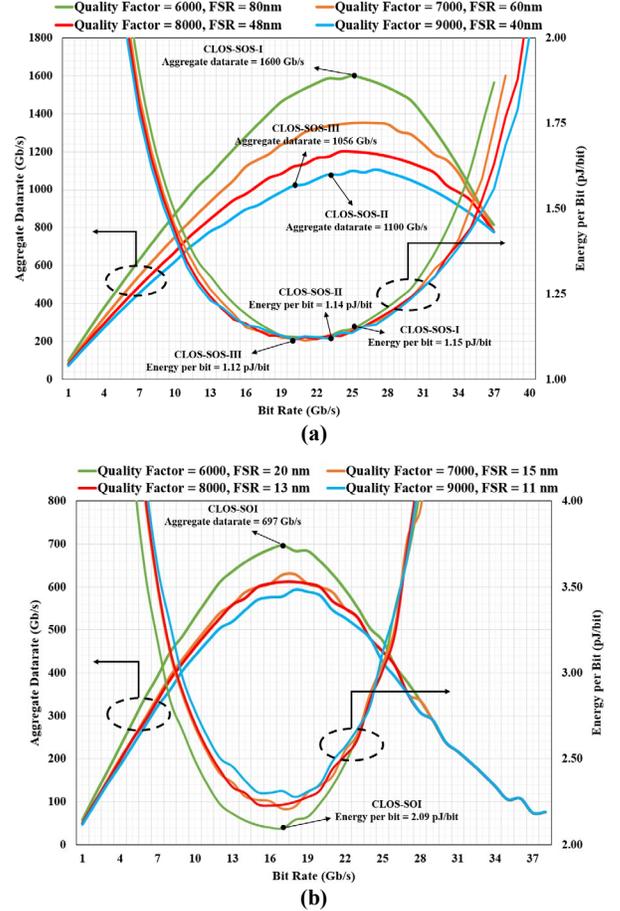

**Fig. 5:** Aggregate data rate and total energy-per-bit (EPB) values for (a) *SOSPh* links, and (b) *SOIPh* links, for different Q-factor, FSR, and MAOP value combinations from Table 2. The optical losses, laser efficiency, and other device parameters for this analysis are taken from [12] and [16].

To understand the reason behind this outcome, we extract total four link designs from Fig. 5(a) and 5(b), and list the relevant parameter values for these link designs in Table 3. We also present, in Fig. 6, how the OPB is utilized for the specific *SOSPh* and *SOIPh* link designs from Table 3. From Fig. 6, it is evident that lower losses and higher MAOP for CLOS-SOS-I, CLOS-SOS-II, and CLOS-SOS-III link designs yield greater aggregate data rate and lower EPB values for them, compared to the CLOS-SOI link design. However, note that CLOS-SOS-I, CLOS-SOS-II, and CLOS-SOS-III link designs still do not achieve sub-pJ EPB values as desired. Nevertheless, as the per-wavelength (per-λ) power requirements for the *SOSPh* link designs from Table 3 are far lower than their saturation point (i.e., 5dBm [18]), *these SOSPh link designs still have potential to achieve better (<1pJ/bit) EPB values by simply allowing greater than 22dBm MAOP per link*. Thus, from these results, we can conclude that *our proposed SOSPh device platform can pave the way for realizing ultra-low-energy on-chip interconnects of the future.*

Excellent link-level results for *SOSPh* platform cannot guarantee good performance at the system-level, especially for the real-world traffic scenarios of on-chip communication. Therefore, to establish a clear winner between the *SOIPh* and *SOSPh* platforms, we present benchmark-driven system-level analysis in the next section.

Table 3: $N_\lambda$ and bit-rate for different variants of CLOS PNoC.

| Extracted Link Designs | $N_\lambda$ | Bit-Rate (Gb/s) | Q-Factor | FSR (nm) | Power Per-$\lambda$ (dBm) |
|---|---|---|---|---|---|
| CLOS-SOI | 41 | 17 | 6000 | 18 | 1.31 |
| CLOS-SOS-I | 64 | 25 | 6000 | 80 | -8.08 |
| CLOS-SOS-II | 44 | 25 | 9000 | 40 | -4.83 |
| CLOS-SOS-III | 48 | 22 | 9000 | 40 | -3.16 |

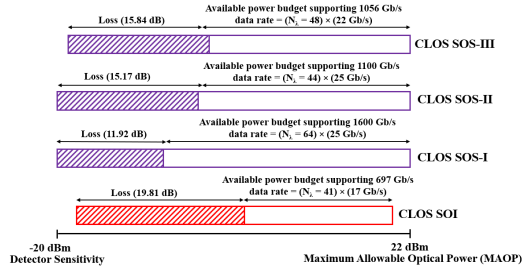

**Fig. 6:** Distribution of optical power budget (OPB) for different *SOIPh* and *SOSPh* link designs from Table 3.

## 5. SYSTEM-LEVEL EVALUATION

### 5.1 Evaluation Setup

We have done our evaluation on a 256-core system implementing 8-ary 3-stage CLOS topology PNoC [33]. The system has 8 clusters (C1-C8) with 32 cores in each cluster, a group of four cores are connected to a concentrator inside a cluster. There are 8 concentrators in each cluster, and an electrical router connected to them to realize inter-concentrator communication. Point-to-point photonic links are used for inter-cluster communication; a total of 56 single-waveguide links are used to connect all 8 clusters of the CLOS PNoC. Depending on the physical location of source and destination, the point-to-point photonic links use forward or backward propagating wavelengths. Two laser sources are used to enable forward and backward communication in PNoC. The CLOS PNoC uses 1X2, 1X7, and 1X4 splitters to power the 56 waveguides.

We performed benchmark-driven simulation-based analysis to evaluate the impact of *SOSPh* and *SOIPh* links from Table 3 on the performance and energy-efficiency of CLOS PNoC architecture. We used $N_\lambda$ and bit-rate values from Table 3 to model four variants of CLOS PNoC using a cycle-accurate NoC simulator. We evaluated performance for a 256-core single-chip architecture at a 22nm CMOS node. We kept the number of WGs and basic floorplan of the architectures constant across all the variants. We used real-world traffic from applications in the PARSEC benchmark suite [13]. GEM5 full system simulation [14] of parallelized PARSEC applications was used to generate traces that were fed into our cycle-accurate NoC simulator. In GEM5 simulations, we set a "warmup" period of 100 million instructions and then captured traces for the subsequent 1 billion instructions. In our benchmark-driven simulations, we evaluated average packet latency, and energy-per-bit (EPB) values for different variants of CLOS PNoC.

### 5.2 Experimental Results

Fig. 7(a) represents a comparison of average packet latency values for the CLOS-SOI, CLOS-SOS-I, CLOS-SOS-II and CLOS-SOS-III PNoCs. As evident, compared to CLOS-SOI PNoC, SOS based PNOCs CLOS-SOS-I, CLOS-SOS-II and CLOS-SOS-III, respectively, have 45%, 26% and 26% lower average packet latency on average. From Table 3, CLOS-SOS variants have higher $N_\lambda$ than CLOS-SOI PNoC, which increases the number of concurrent bits transferred over the network for the CLOS-SOS variants, reducing their average packet latency values. As CLOS-SOS-I has the highest $N_\lambda$, it has the least average packet latency. In addition to higher $N_\lambda$, SOS variants also have better bit-rate, which increases the rate at which the bits are transferred, eventually contributing to the reduced latency. We can observe that CLOS-SOS-II and CLOS-SOS-III achieve same average latency, this is because CLOS-SOS-II has higher bit-rate which is compensated by CLOS-SOS-III's higher $N_\lambda$.

As evident from Fig. 7(b), CLOS-SOS-I, CLOS-SOS-II and CLOS-SOS-III have 29%, 37% and 36% lower EPB compared CLOS-SOI on average. As the average latency for the SOS variants is less than CLOS-SOI, energy dissipated is also less. The EPB of CLOS-SOS-I is greater than CLOS-SOS-II and CLOS-SOS-II, as greater $N_\lambda$ leads to increase in the number of MR modulators and MR detectors in CLOS-SOS-I, which in turn increases the total energy consumption.

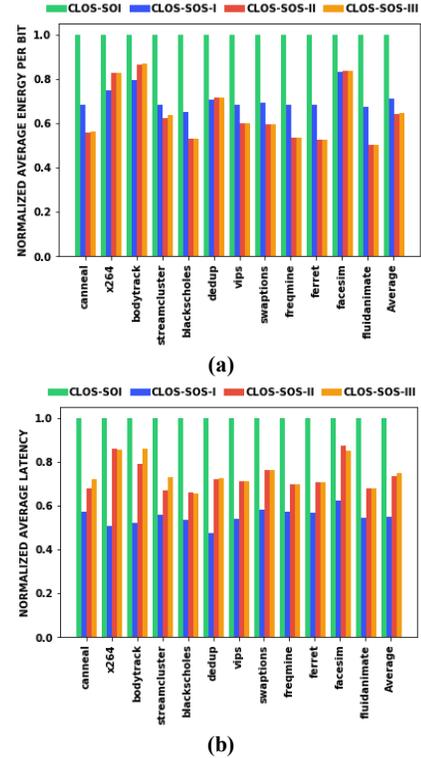

**Fig. 7** (a) Average packet latency, and (b) energy-per-bit (EPB) comparisons for different variants of CLOS PNoC across PARSEC benchmarks. All results are normalized to the baseline CLOS-SOI PNoC results.

In summary, we showed that PNoCs that are implemented using our proposed *SOSPh* devices and links are more energy-efficient, and achieve lower latency values, compared to the PNoCs implemented using the conventional *SOIPh* devices and links. These results corroborate the capabilities of our proposed *SOSPh* platform based PNoCs to achieve high-bandwidth data transfers with greater energy efficiency and lower latency compared to *SOIPh* platform.

## 6. RELATED WORK

Significant research work (e.g., [16][17][23][29]) is available in the literature that focuses on characterizing the two-photon absorption (TPA) and other types of optical non-linear effects in silicon waveguides and resonators. For example, [16] and [17] describe how TPA induced FCD and FCA effects in silicon limit the MAOP in *SOIPh* links, restricting the scalability of their aggregate data rate and energy-efficiency. However, no prior work has yet explored a solution to the TPA-induced scalability shortcomings of *SOIPh* in-terconnects. We for the first time presented SOS-based device platform as a potential solution to the TPA-related scalability issues in on-chip photonic interconnects.

Several SOS-based photonic devices have already been prototyped to be operated near 4μm wavelength. These prototypes include on-chip quantum cascade laser sources (e.g., [26]), photonic waveguides and MRs (e.g., [28]-[30]), grating couplers (e.g., [27]). Information obtained from all these prototype works, when combined with the knowledge base from this paper, can catalyze cross-layer research in the area of *SOSPh* interconnects design, which can enable the widespread adoption of *SOSPh* platform for realizing extreme-scale on-chip and off-chip communication architectures.

## 7. OVERHEADS AND CHALLENGES

To compare the footprint area of SOS and SOI variants of CLOS PNoC architecture from Table 3, *SOIPh* MR has footprint area of 78μm$^2$, whereas the footprint areas for SOS-I, SOS-II and SOS-III MRs are 177 μm$^2$, 707 μm$^2$ and 708 μm$^2$ respectively. The footprint area of a 1cm long rectilinear *SOIPh* waveguide is 4500μm$^2$, whereas the footprint area of 1cm long rectilinear *SOSPh* waveguide is 9700μm$^2$. In terms of CLOS PNoC architecture, the total footprint area for SOI-based CLOS PNoC architecture is 0.4 mm$^2$, whereas the footprint area for SOS-I, SOS-II, and SOS-III based CLOS PNoC architectures are 3.1 mm$^2$, 2.3 mm$^2$ and 2.4 mm$^2$, respectively. This comparison clearly shows that SOS links and PNoCs have higher footprint area compared to SOI links and PNoCs.

To support our *SOSPh* platform for realizing communication architectures, heterogeneity needs to be introduced in the already established hierarchy of interconnection networks. Traditional fiber optics systems for inter-cluster, inter-datacenter, and long haul networks still running on O, L and C optical bands. In contrast, *SOSPh* platform operates with wavelengths between 2.5μm-4μm. Therefore, additional specialized equipment and support are needed to introduce *SOSPh* interconnects in this established hierarchy, which is likely to incur extra cost. Nevertheless, it is worth bearing this extra cost, especially considering the energy and performance benefits of *SOSPh* platform shown in this paper.

## 8. CONCLUSIONS

Conventional SOI-based photonic interconnects have limited bandwidth-energy scalability due to the optical non-linear effects in silicon, especially the two-photon-absorption (TPA) effect. In this paper, we presented silicon-on-sapphire (SOS) device platform as a solution to the scalability limitations of SOI-based interconnects. We developed new compact models for SOS devices, utilizing which we formulated new guidelines for designing SOS links and PNoCs. Our link-level analysis showed that SOS links can achieve aggregate data rate of >1Tb/s, which is significantly better than SOI links. Our system-level analysis with CLOS PNoC architecture showed that PNoCs that are designed using SOS devices and links can achieve up to 45% lower latency and 37% lower EPB compared to the PNoCs implemented using the conventional SOI devices and links. These promising results prove that SOS-based PNoCs can achieve high-bandwidth data transfers with low latency and greater energy-efficiency, compared to the traditional SOI-based PNoCs.